\documentclass{ws-procs9x6-cpt22}
\begin{document}

\newcommand{\refeq}[1]{(\ref{#1})}
\def\etal {{\it et al.}}

\title{Search for CPT- and Lorentz-Invariance Violation in the \\
Muon g-2 Experiment at Fermilab\\
}

\author{B.\ Mitra}

\address{Department of Physics and Astronomy, University of Mississippi,\\
University, MS 38677, USA}

\author{On behalf of the Muon g-2 Collaboration}

\begin{abstract}
Muon g-2 data can be used to study sidereal variations in the anomalous muon precession rate, which is one of the important signatures of CPT- and Lorentz invariance violation. The discussion in this work will focus on the framework that has been used to study sidereal variations in the anomalous muon precession rate in Muon g-2 Run 2 data. Also, a brief introduction will be given about the blinding framework that will be used on Run 2/3 data of the Muon g-2 experiment. This blinding framework will keep the actual result of the experiment masked from the analyzer during the analysis process. 
\end{abstract}

\bodymatter

\section{CPT- and Lorentz-violating signatures in Muon g-2 }

The Standard-Model Extension(SME) is a framework that describes CPT- and
Lorentz-invariance violation by adding new terms to the Standard-Model
Lagrangian. The minimal SME Lagrangian$^1$ for muon is given by:
\begin{equation}
\begin{split}
\mathcal{L}=\ -a_k\bar{\psi}\gamma^{k}\psi\ -b_k\bar{\psi}\gamma_5\gamma^{k}\psi
\ -\frac{1}{2} H_{k\lambda}\bar{\psi}\sigma^{k\lambda}\psi \\
\ +\frac{1}{2} i c_{k\lambda}\bar{\psi}\gamma^{k}{{\overset\leftrightarrow D}^{\lambda}}\psi
\ +\frac{1}{2} id_{k\lambda}\bar{\psi}\gamma_5\gamma^{k}{\overset\leftrightarrow D}^{\lambda}\psi
\end{split}
\label{aba:appeq1}
\end{equation}
In the above equation, all terms violate Lorentz invariance. The coefficients $a_{k}$
and $b_{k}$ are CPT odd and other coefficients in Eq. (1) are CPT even. 
Sidereal and annual variations in the anomalous precession frequency of the muon, i.e., $\omega_{a}$
can be used to determine the transverse component of $b_{k}$. Difference in $\omega_{a}$
between $\mu_{+}$ and $\mu_{-}$ data can be used to determine $b_{z}$, $H_{XY}$, $d_{ZO}$.

\section{Measurement of transverse component of $b_{k}$}
In the Muon g-2 experiment instead of $\omega_{a}$, the ratio of $\omega_{a}$ and
$\widetilde{\omega_{p^{'}}}$ is measured. Where $\widetilde{\omega_{p^{'}}}$ is the
proton's Larmor-precession frequency in a water sample mapping the magnetic field and weighted 
by the muon distribution. The ration of $\omega_{a}$ and
$\widetilde{\omega_{p^{'}}}$ is measured instead of only $\omega_{a}$ to nullify the
effects of variations of the magnetic field on the measurement.

\section{Lomb-Scargle search for sidereal oscillations in  $\omega_{a}$ }
In the Lomb-Scargle method, the spectral power $P_{N}(\omega)$ is calculated corresponding to each
frequency $\omega$. This $P_{N}(\omega)$ is a measure of the statistical significance, or 
likelihood, of a signal at a given frequency. A Higher value of $P_{N}(\omega)$ implies a more 
significant periodic signal at that frequency $\omega$.

\section{Multi-parameter fit search for sidereal oscillations in $\omega_{a}$ }
This procedure is based on a four-parameter fit model. In the model, following equation is used:
\begin{equation}
\mathcal{R}(t)=C_0 + A_0cos(\frac{2\pi t}{T_0} + \phi_0)
\label{aba:appeq2}
\end{equation}
In Eq. (2) $C_0$ is time average of $R$, which is a constant in time. $A_0$ is oscillation
amplitude. This method is appropriate for many kinds of data.
\section{Run by Run analysis  of $\omega_{a}$ data }
In the Muon g-2 experiment, each run is approximately one hour long. In this method, the value of 
$\omega_{a}$ is extracted for each run. This gives an unevenly sampled data. Both the Lomb-Scargle and
multi-parameter fit method is used on this unevenly sampled extracted data.
\section{Folded analysis of $\omega_{a}$ data }
In this method, data is folded over evenly sampled time bins. Five different window size have been
used in this analysis. Lomb-Scargle, multi-parameter fit, and FFT methods are then applied to
the extracted $\omega_{a}$ time-series data.

\section{Blinding framework on g-2 CPTLV analysis}
This method will be used on the Run2/3 data set analysis. At first fake signals will be injected at random frequencies and with random amplitudes. The total number of injected fake signals will be
blinded. The amplitude of the fake signals will be more than 2 ppm and less than 10 ppm.

\section*{Acknowledgments}
This work was supported in part by the US DOE and Fermilab.

\end{document}